\documentclass[aps,
pre,
preprint,
superscriptaddress,
showpacs,
amsmath
]{revtex4-1}

\usepackage{amsfonts,amsmath,amssymb}
\usepackage{graphicx}
\usepackage{url}
\usepackage{latexsym}
\usepackage{theorem}
\usepackage{psfrag}

\newcommand{\be}{\begin{equation}}
\newcommand{\ee}{\end{equation}}

\newcommand{\e}{\varepsilon}
\newcommand{\g}{\gamma}
\newcommand{\vp}{\varphi}
\newcommand{\w}{\omega}

\def \ii {\text{i}}

\allowdisplaybreaks
\begin{document}

\title{Synchronization transitions in ensembles of noisy oscillators with
bi-harmonic coupling}
\author{Vladimir Vlasov}
\affiliation{Institute for Physics and Astronomy, University of Potsdam,
 14476 Potsdam, Germany}
\author{Maxim Komarov}
\affiliation{Institute for Physics and Astronomy, University of Potsdam,
 14476 Potsdam, Germany}
\affiliation{Department of Control Theory, Nizhni Novgorod State University,
Gagarin Av. 23, 606950,
Nizhni Novgorod, Russia
}\author{Arkady~Pikovsky}
\affiliation{Institute for Physics and Astronomy, University of Potsdam,
 14476 Potsdam, Germany}
\affiliation{Department of Control Theory, Nizhni Novgorod State University,
Gagarin Av. 23, 606950,
Nizhni Novgorod, Russia
}

\date{\today}

\begin{abstract}
We describe synchronization transitions in an ensemble of globally coupled
phase oscillators with a bi-harmonic coupling function, and two sources of disorder - diversity
of intrinsic oscillatory frequencies and external independent noise. 
Based on the self-consistent formulation,
we derive analytic solutions for different synchronous states. 
We report on various non-trivial transitions from incoherence to synchrony where possible scenarios include: 
simple supercritical transition (similar to classical Kuramoto model), subcritical transition with large area of bistability of incoherent and synchronous solutions, and also appearance of symmetric two-cluster solution which can coexist with regular synchronous state.
Remarkably, we show that the interplay between relatively small white noise and finite-size fluctuations can lead to metastable asynchronous solution.
\end{abstract}
 \maketitle

 \section{Introduction}

	In the theory of synchronization, Kuramoto model of globally
coupled phase oscillators \cite{Kuramoto-75,Kuramoto-84,Acebron-etal-05} is one
of the most popular setups for describing synchronization phenomena. The case of
a harmonic sin-coupling function is well studied in
literature~\cite{Kuramoto-84,Ott-Antonsen-08,Ott-Antonsen-09}. 
However, when reducing complex nonlinear oscillatory dynamics to relatively
simple phase models, one often has to deal with 
multiharmonic coupling functions~\cite{Kuramoto-84}.
In this paper we concentrate on the particular case of the Kuramoto model of
globally coupled phase oscillators with a bi-harmonic interaction function.
Such type of interaction between oscillators arises in several realistic
physical setups including: (i) classical Huygens' setup with pendulum clocks
suspended on a common beam with both vertical and horizontal
displacements~\cite{Czolczynski_etal-13,Komarov-Pikovsky-14}; (ii) recently
experimentally observed $\varphi$--Josephson junctions, where the dynamics of a
single junction in the array is governed by a double-well energy
potential~\cite{Goldobin_etal-11}; (iii) globally coupled electrochemical
oscillators~\cite{Kiss-Zhai-Hudson-05,Kiss-Zhai-Hudson-06}, where the second
harmonics has been obtained from the experimental data.

In our work we take into account two main sources of disorder that hinder
synchronization, namely diversity of oscillators' frequencies, and white
additive noise acting on the phases. 
	Recent theoretical studies indicate that in the noise-free case the
Kuramoto model with bi-harmonic coupling function is characterized by a variety
of multi-branch entrainment modes \cite{Daido-95, Daido-96a, Winfree-80}.
The latter manifests itself as a multiplicity (multistability) of different
synchronous states, with distinct redistributions of oscillators between two
stable branches of microscopic dynamics
\cite{Komarov-Pikovsky-13a,Komarov-Pikovsky-14}.
In this work we show that the action of white noise removes the multiplicity,
however, the overall picture of transitions from incoherence to synchrony is
non-trivial and can be quite complex in
comparison to the standard Kuramoto model.
		
	The paper is organized as follows. First, we formulate the problem and
perform a special, suitable for the analysis,
parametrization of the system. Then, we present a self-consistent approach
allowing us to find order parameters in dependence on the coupling constants and
disorder in an analytic form. Wherever possible, we perform the stability
analysis. And at last, we discuss the limiting case of small noise intensity and
its relation to the noiseless situation. In conclusion, we summarize and collect
all the findings together.

\section{Basic Model}
In this paper we study an ensemble of phase oscillators (phase variables
$\phi_k$), 
subject to a mean-field
 bi-harmonic coupling and noise [in this formulation we use ``primed''
variables, which will be in short transformed to dimensionless ones]:
	\be
		\label{bi-Kur.1}
		\frac{d \phi_k}{dt'}=\w_k\Delta'+{\e\over N}\sum_{j=1}^N
\sin(\phi_j-\phi_k)+{\g\over N}\sum_{j=1}^N \sin(2\phi_j-2\phi_k) +
\sqrt{D'\,}\xi_k(t')\;.
	\ee
Here $\w_k$ are normalized natural frequencies of oscillators that are 
assumed to have a symmetrical distribution $g(\w)$ with unit width 
and zero mean value (the latter condition is not a restriction, as it can be
always achieved by transforming to
a properly rotating reference frame). Parameter $\Delta'$ measures the spread of
the distribution
of natural frequencies. Gaussian white noise is defined according to $\langle
\xi_k(t_1')\xi_j(t_2')\rangle=2\delta(t_1'-t_2')\delta_{kj}$.
Parameters $\e$ and $\g$ define the strength of the coupling on the first and
the second harmonics, respectively.
	
Eq.~(\ref{bi-Kur.1}) can be rewritten as
	\be
		\label{bi-Kur.2}
		\frac{d \phi_k}{dt'}=\omega_k\Delta'+\e R_1
\sin(\Theta_1-\phi_k)+\gamma R_2 \sin(\Theta_2-2\phi_k)+\sqrt{D'\,}\xi_k(t'),
	\ee
	where $R_m e^{\ii\Theta_m}=N^{-1}\sum_j e^{\ii m\phi_j}$, $m=1,2$, are
the two relevant order parameters.

	Eq.~(\ref{bi-Kur.2}) has 4 parameters, all of them of dimension $1/t'$:
$\Delta',\e,\gamma,D'$.
	By rescaling time, one parameter can be set to one. We choose to set the
overall coupling $\e+\gamma$ to unity.
	Thus, we introduce $t=(\e+\gamma)t'$ and get 
	\be
		\label{bi-Kur.newnorm.1}
		\dot\phi_k=\omega_k\Delta+q R_1 \sin(\Theta_1-\phi_k)+(1-q) R_2
\sin(\Theta_2-2\phi_k)+\sqrt{D\,}\xi(t),
	\ee
	where $\Delta=\Delta'/(\e+\gamma)$, $q=\e/(\e+\gamma)$ and
$D=D'/(\e+\gamma)$.
	Parameter $q$ describes relation between coupling coefficients $\e$ and
$\g$,
	such that the case $q=0$ corresponds to a pure second harmonic coupling
with $\e=0$,
	while $q=1$ corresponds to a pure Kuramoto-type first harmonic coupling
with $\g=0$.
	In this new normalization, increasing or decreasing of coupling
strength 
	is equivalent to decreasing or increasing of the disorder parameters
$\Delta$ (spread of frequencies) and $D$ (noise), 
	while keeping a constant relation $\Delta/D$ between them.
	This suggests to introduce new parameters $T,s$ in such a way that
$\Delta=(1-s)T$, $D=sT$. 
	Therefore, parameter $T$ measures the overall disorder (noise plus
spread of frequencies), normalized by the overall coupling strength $\e+\g$.
Parameter $s$ measures the relation between  parameters $\Delta$ (width of
frequency distribution)
	and $D$ (noise): for $s=0$ the system is purely deterministic, while for
$s=1$ it describes
	an ensemble of identical noisy oscillators.  
	Summarizing, Eq.~(\ref{bi-Kur.newnorm.1}) with new parameters $q,T,s$
becomes
	\be
		\label{bi-Kur.newnorm.2}
		\dot\phi_k=\omega_k(1-s)T+q R_1 \sin(\Theta_1-\phi_k)+(1-q) R_2
\sin(\Theta_2-2\phi_k)+\sqrt{sT\,}\xi_k(t).
	\ee

	We consider the thermodynamic limit $N\rightarrow\infty$, then the order
parameters are just ensemble averages 
	$R_m e^{\ii\Theta_m}=\langle e^{\ii m\phi}\rangle$
 and in the thermodyanmical limit can be represented through the conditional probability density
function of 
	the phases $\rho(\vp | \omega)$, as
	\be
		\label{R_m.1}
		R_{m}e^{\ii\Theta_m}=\langle e^{\ii m\vp}\rangle=
\int \int  g(\omega)\,\rho(\vp |\omega)e^{\ii m\vp}d\vp \,d\omega
	\ee
	
In thermodynamic formulation we use the variable $\vp$ to describe the phase and skip all indices, therefore, according to (\ref{bi-Kur.newnorm.2}) the equation for the phase variable $\vp$ at given $\w$ reads
	\be
		\label{bi-Kur.3}
		\frac{d \vp}{dt}=\omega(1-s)T+q R_1 \sin(\Theta_1-\vp)+(1-q)
R_2 \sin(\Theta_2-2\vp)+\sqrt{sT\,}\xi(t)\;.
	\ee
	The Fokker-Planck equation for $\rho(\vp,t\,|\omega)$ follows from
Eq.~(\ref{bi-Kur.3}):
	\be
		\label{FP.1}
		\frac{\partial \rho}{\partial t}+\frac{\partial}{\partial
\vp}\Bigl[\Bigl(\w(1-s)T+qR_1\sin(\Theta_1-\vp)+(1-q)R_2\sin
(\Theta_2-2\vp)\Bigr)\rho\Bigr]=
sT\frac{\partial^2\rho}{\partial\vp^2}\,.
	\ee
	
 The limiting noise-free case when $s=0$  has been described in details
in~\cite{Komarov-Pikovsky-13a,Komarov-Pikovsky-14}.
In this paper we will present a general analysis for systems with noise and a
finite distribution
of frequencies. We will treat the limit  $s\ll 1$, which appears to be singular,
separately.
The other limiting case $s=1$ is the case of the identical natural frequencies,
in terms of the analysis presented below it is not special. However, 
for $s=1$ an additional stability analysis can be performed, thus this case will
be also considered in details separately.


\section{Stationary solutions: Self-consistent approach}
A disordered state with a uniform distribution of phases $\rho=(2\pi)^{-1}$ and
vanishing order
parameters $R_1=R_2=0$ is
always a solution of the system (\ref{R_m.1},\ref{FP.1}). Additionally, we
expect nontrivial synchronized states
of two types: (i) all order parameters are non-zero, and (ii) a symmetric
2-cluster distribution
where all odd order parameters vanish $R_{2m+1}=0$ and $R_{2m}\neq 0$,
$m \in \mathbb{N}_0$. 

\subsection{Solution in a parametric form}

Due to the symmetry of the coupling function and of the distribution of natural
frequencies $g(\w)$,
the nontrivial solutions are stationary states (what means that the frequency of
the mean fields is exactly
the average oscillator frequency)  with $\Theta_1=\Theta_2=0$.
Thus, the stationary conditional probability density function
$\rho(\vp\,|\,\omega)$ satisfies
the stationary Fokker-Planck equation
	\be
		\label{FP.st.1}
		\frac{\partial}{\partial
\vp}\Biggl[\Biggl(\w(1-s)-\frac{qR_1}{T}\sin(\vp)-\frac{(1-q)R_2}{T}\sin
(2\vp)\Biggr)\rho\Biggr]=
s\frac{\partial^2\rho}{\partial\vp^2}\,,
	\ee
	where because of symmetry
	\be
		\label{R_m.st.1}
		R_{m}=\int \int  g(\omega)\,\rho(\vp\,|\,\omega)\cos( m\vp)d\vp
\,d\omega
	\ee

To find the solutions of this self-consistent system explicitly,	it is
convenient to introduce
two new auxiliary variables $R$ and $\alpha$ (together with definitions $u$, $v$, and $x$) according to
	\be
		\label{notation.1}
		\begin{split}
			R&=\sqrt{\Bigl(q R_1/T\Bigr)^2+\Bigl((1-q)
R_2/T\Bigr)^2\,}, \\
			u&=\cos\alpha=\frac{q R_1}{TR},\\
			 v&=\sin\alpha=\frac{(1-q) R_2}{TR},\\
			 x&=\frac{\w}{R}
		\end{split}
	\ee
	Then the stationary Fokker-Plank equation~(\ref{FP.st.1}) for the
stationary distribution density 
	$\rho(\vp\,|\, x)$ (which depends on $R,\alpha,s$
as parameters) reads
	\be
		\label{FP.st.2}
		\frac{\partial}{\partial
\vp}\Bigl[R\Bigl(x(1-s)-u\sin(\vp)-v\sin(
2\vp)\Bigr)\rho\Bigr]=s\frac{\partial^2\rho}{\partial\vp^2}\,.
	\ee
	An explicit solution of \eqref{FP.st.2}
	can be written as double integrals, but
	practically it is more convenient to solve it in the Fourier modes
representation
	\be	
		\label{Four.1}
		\rho(\vp\,|\,x)=\frac{1}{2\pi}\sum_n C_n(\alpha,R,
s,x)e^{\ii n\vp},\quad C_n(\alpha,R, s,x)=\int_0^{2\pi}\rho e^{-\ii
n\vp}d\vp,\quad C_0=1.
	\ee
	Substituting~(\ref{Four.1}) in Eq.~(\ref{FP.st.2}) we obtain
	\be
		\label{FP.st.Four.1}
		\begin{split}
			0&=\int_0^{2\pi}\Biggl[-\frac{\partial}{\partial
\vp}\Bigl[R\Bigl(x(1-s)-u\sin(\vp)-v\sin(
2\vp)\Bigr)\rho\Bigr]+s\frac{\partial^2\rho}{\partial\vp^2}\Biggr] e^{-\ii
k\vp}\,d\vp=\\
			&=R\Biggl[\Bigl(-\ii x(1-s)k-k^2s/R\Bigr)C_k+\ii
ku\frac{C_{k-1}-C_{k+1}}{2\ii}+\ii kv\frac{C_{k-2}-C_{k+2}}{2\ii}\Biggr].
		\end{split}
	\ee
	Thus, from~(\ref{FP.st.Four.1}) we obtain a system of linear algebraic
equations for the mode amplitudes:
	\be
		\label{FP.st.Four.2}
		2\Bigl(\ii
x(1-s)+ks/R\Bigr)C_k+u(C_{k+1}-C_{k-1})+v(C_{k+2}-C_{k-2})=0\;,
	\ee
	which can be solved by standard methods after a proper truncation to a
finite number 
	of Fourier modes (which controls accuracy of the solution)
	is performed.
	After finding $C_{1,2}(\alpha,R, s,x)$, we have to calculate integrals
	\be
		\label{st.F.1}
		F_{1,2}(\alpha,R, s)=\int g(Rx)\,{\rm Re}\left[C_{1,2}(\alpha,R,
s,x)\right]dx.
	\ee
This allows us to represent the order parameters as
	\be
		\label{st.R_m.1}
		R_{1,2}(\alpha,R, s)=R\int g(Rx)\,{\rm
Re}\left[C_{1,2}(\alpha,R, s,x)\right]dx=RF_{1,2}(\alpha,R, s).
	\ee
	Substituting this in Eq.~\eqref{notation.1}, we obtain our parameters
$T,q$ as functions of the auxiliary
	variables
	\be
		\label{sol.st.3}
		\begin{split}		
T&=\frac{1}{\frac{\cos\alpha}{F_1}+\frac{\sin\alpha}{F_2}},\\		
q&=\frac{\frac{\cos\alpha}{F_1}}{\frac{\cos\alpha}{F_1}+\frac{\sin\alpha}{F_2}}
=\frac{1}{1+\frac{F_1}{F_2}\tan\alpha}.
		\end{split}
	\ee
Summarizing, we have obtained the explicit solution of the self-consistent
problem:
for each fixed $s$, by  varying $\alpha\in[0,\pi/2]$ and $R\in[0,\infty)$, we 
obtain the solution in a parametric form: $R_{1,2}=R_{1,2}(\alpha,R)$ according
to \eqref{st.R_m.1},
$T=T(\alpha,R)$ and $q=q(\alpha,R)$ according to \eqref{sol.st.3}. 

The case of purely two-cluster state with $R_1=0$ corresponds to $\alpha=\pi/2$,
it is singular in (\ref{st.R_m.1},\ref{sol.st.3}).
Here the solution is represented as
\be
R_2=R F_2,\qquad T=(1-q) F_2.
\label{twocl}
\ee
	
	Thus, by the method presented above, it is possible to find stationary
solutions of the Eq.~(\ref{FP.1}) for any given $q,T$ and $s$. In the general
case of $s<1$,  Eq.~(\ref{FP.1}) is integro-differential equation and the
analysis of the stability 
of all solutions is quite difficult, except for the simplest incoherent solution
$\rho=(2\pi)^{-1}$. 
However, in the limiting case of identical natural frequencies $s=1$, density
$\rho$ is $\omega$-independent, and
integration in~(\ref{R_m.st.1}) over frequencies gives always $1$. 
This means that Fourier modes ~\eqref{Four.1}  are in fact the order
parameters: 
$R_{m}(q,T,1)={\rm Re}\left[C_{m}(q,T,1,0)\right]$. The same is valid also
for the full time-dependent problem: it can be written
as a system of nonlinear ordinary differential
equations for  time-depended Fourier modes of the density, which
can be analyzed for stability after a proper truncation. Thus, we start with
this particular case.

\subsection{Limiting case of identical oscillators}

	Here we describe the case of identical natural frequencies, what means
that $\Delta=(1-s)T=0$ or $s=1$.
		As discussed above, in this case also a stability analysis is
possible, which we outline below.
	First, let us represent Eq.~(\ref{FP.1}) in terms of Fourier modes $C_m$
which are related
	to the complex order parameters $C_{m}^*=R_m e^{\ii \Theta_m}$ (the
procedure is the same as in obtaining \eqref{FP.st.Four.1}):
	\be
		\label{FP.id.Four.1}
		\frac{dC_k}{dt}=-k^2TC_k+ kq\frac{C_1C_{k-1}-C_1^*C_{k+1}}{2}+
k(1-q)\frac{C_2C_{k-2}-C_2^*C_{k+2}}{2}.
	\ee
We are interested in stability of
 a stationary solution, obtained according to (\ref{st.R_m.1},\ref{sol.st.3}) or
\eqref{twocl}.

	For the linear stability analysis, a small perturbation around
stationary solution
$\tilde{C}_k$ should be added, 
	thus we set $C_k=\tilde{C}_k+c_k$ in Eq.~(\ref{FP.id.Four.1}), and in
the first order in $c_k$ obtain
	\begin{equation}
		\label{FP.id.Four.stab.2}
		\begin{split}		
\frac{dc_k}{dt}=&-k^2Tc_k+\frac{kq}{2}\Bigl(c_1\tilde{C}_{k-1}-c_1^*\tilde{C}_{
k+1}\Bigr)+\\		
&+\frac{k(1-q)}{2}\Bigl(c_2\tilde{C}_{k-2}-c_2^*\tilde{C}_{k+2}+\tilde{C}_2c_{
k-2}-\tilde{C}_2^*c_{k+2}\Bigr),
		\end{split}
	\end{equation}
	Equation~(\ref{FP.id.Four.stab.2}) is an infinite system, but
	because the amplitudes of modes with large $k$ tend to zero, it is
possible to truncate it at some large $N$, and to 
	write a finite system of Eqs.~(\ref{FP.id.Four.stab.2}), with $k$
varying from $1$ to $N$, where $N$ is large enough. By finding a maximum
eigenvalue of the corresponding matrix for values $q$ and $T$ of interest, 
it is possible to find stability properties of the solution $\tilde{C}_k$ and 
build the boundary $q=q(T)$ where the solution $\tilde{C}_k$ changes its
stability. This can be done both
for general solutions  (\ref{st.R_m.1},\ref{sol.st.3}) and for the  two-cluster
states \eqref{twocl}.
	
\begin{figure}[!htb]		
\centerline{(a)\includegraphics[width=0.49\columnwidth]{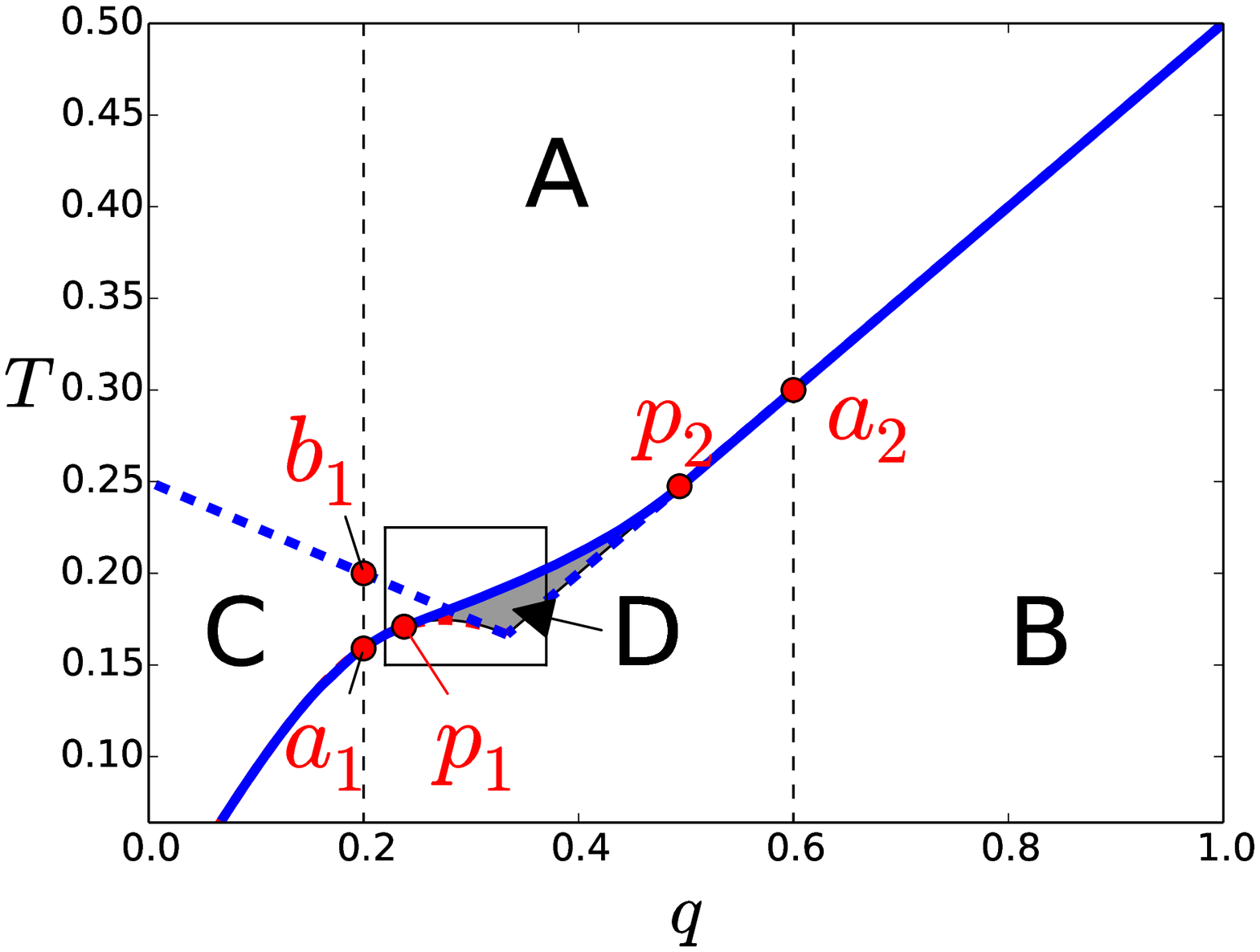}
(b)\includegraphics[width=0.49\columnwidth]{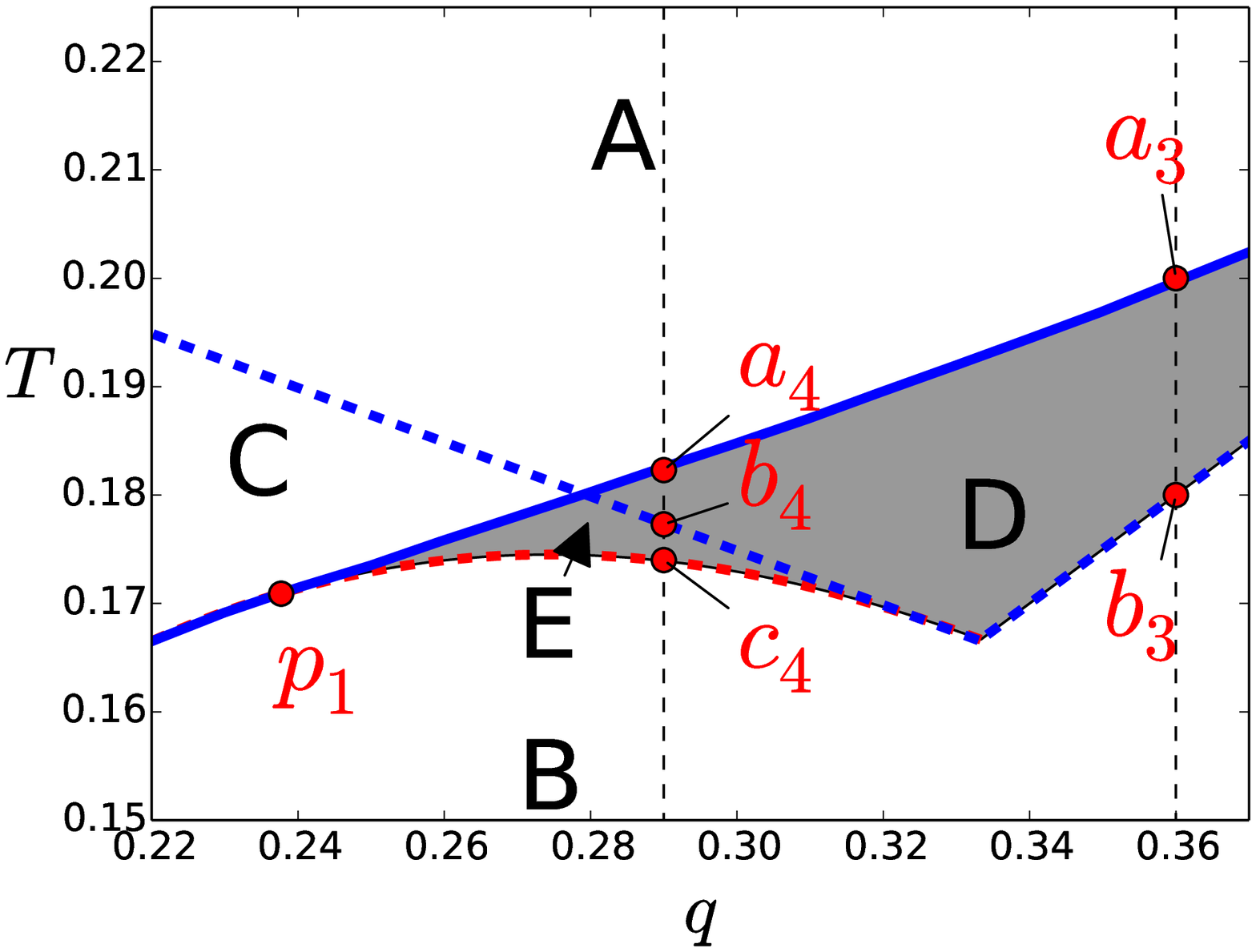}}
		\caption{(a) Different regimes in the parameter plane ($q,T$)
are shown for $s=1$. Area A: asynchronous solution. Area $B$: coherent regime
with $R_{1,2}\neq 0$. Area $C$: two-cluster coherent regime with only order
parameter $R_2 \neq 0$, $R_1=0$. Area D: region of bistability of incoherent and
synchronous solutions. Area E: bistability of the two-cluster state and a state
		with $R_{1,2}\neq 0$. Dashed blue lines are stability lines of
the disordered state,  
		obtained from~(\ref{App:qT.asynch.R1})~and~(\ref{App:qT.asynch.R2}).
Between the points $p_1$ and $p_2$ the transition is hysteretic; 
		dashed red line is the stability line of Area C, obtained
from~(\ref{FP.id.Four.stab.2}), it
		coincides with the line where on the branch existing for small
$T$ the first order parameter
		tends to zero. (b) Enlarged central region of panel (a).
Vertical dashed lines are cuts of the diagram illustrated 
		in Fig.~\ref{fig:r12_s1}.}
		\label{fig:qt_id}
	\end{figure}

\begin{figure}[!htb]		
\centerline{(a)\includegraphics[width=0.49\columnwidth]{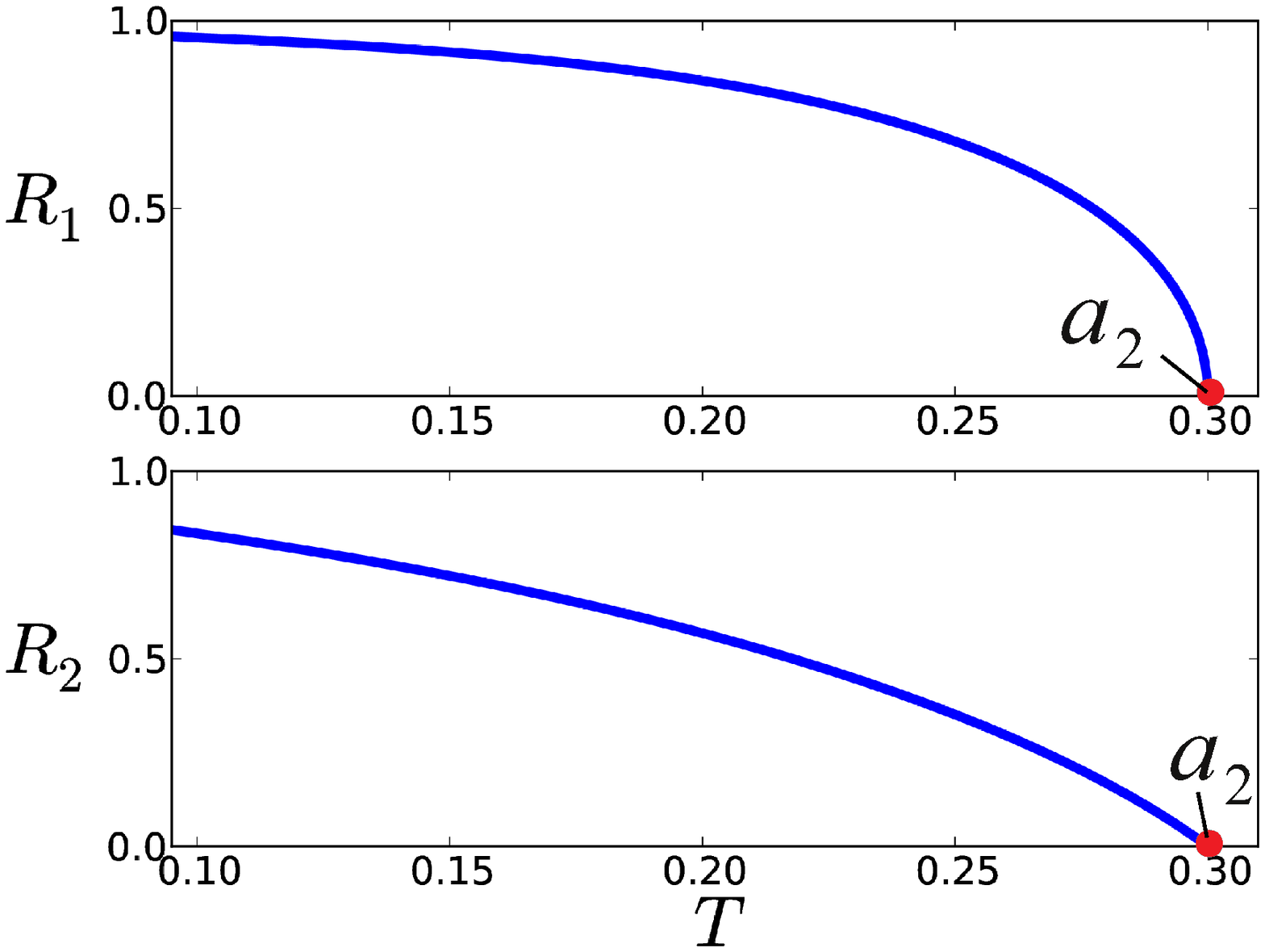}
(b)\includegraphics[width=0.49\columnwidth]{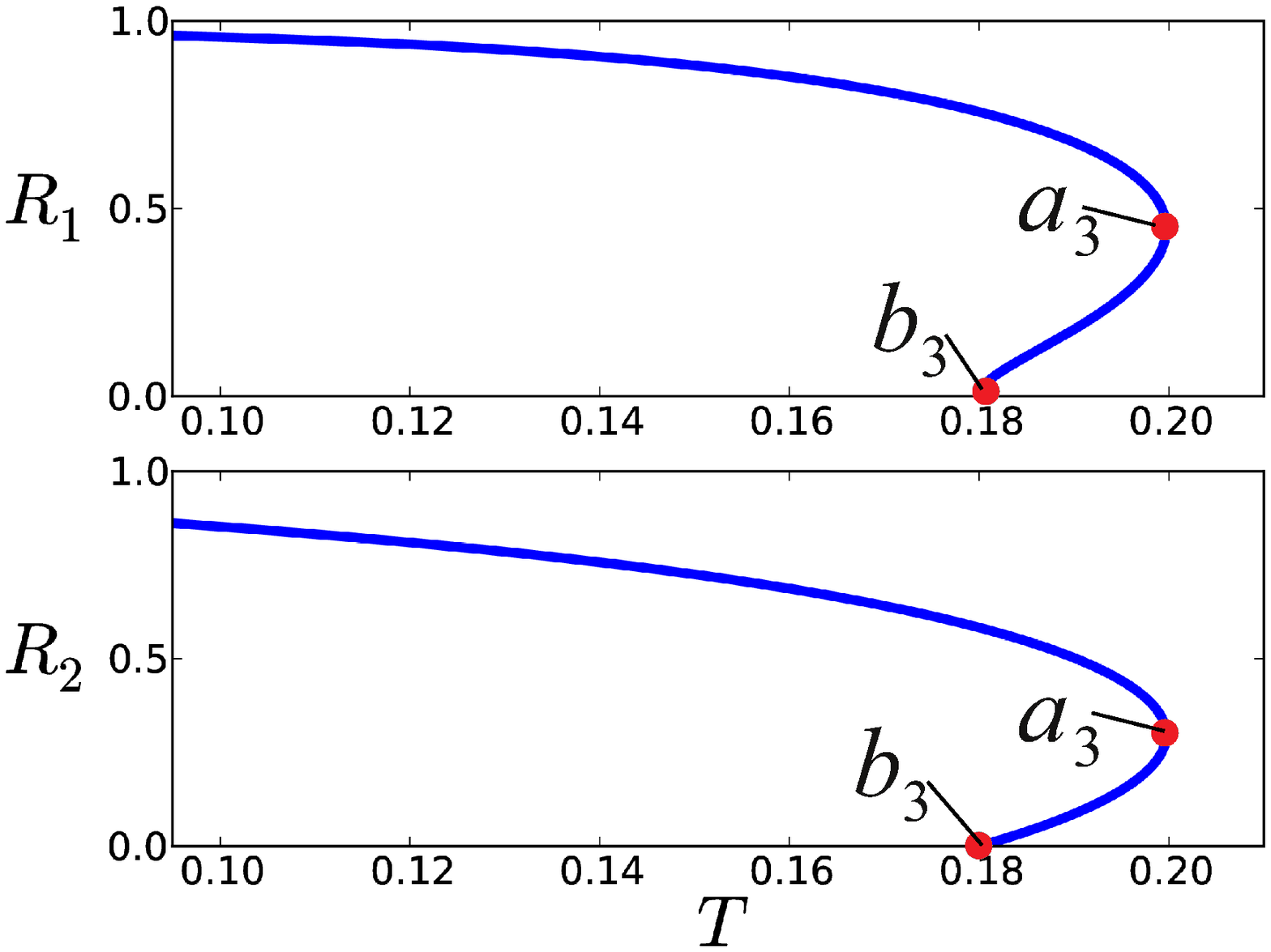}}	
\centerline{(c)\includegraphics[width=0.49\columnwidth]{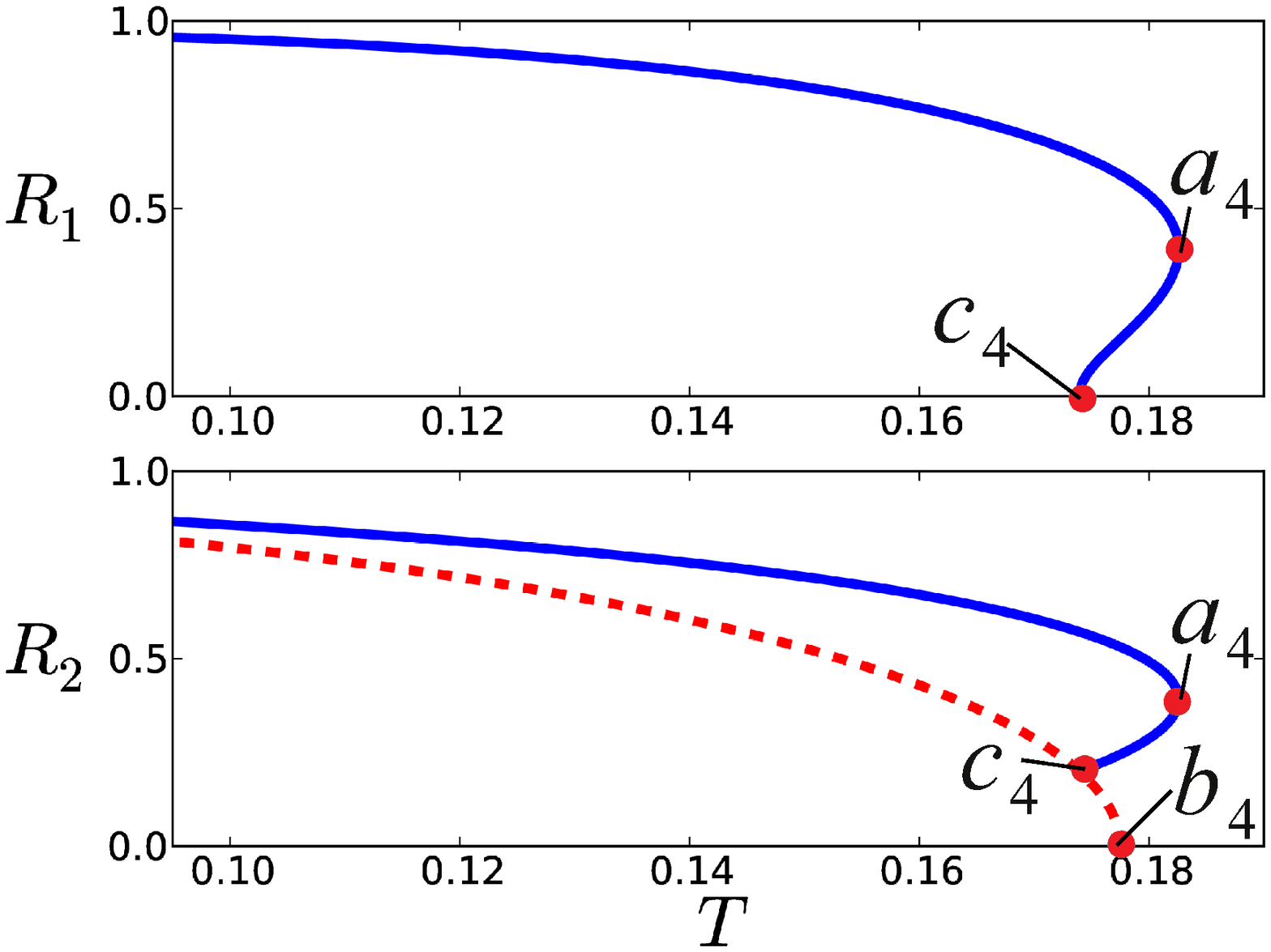}
(d)\includegraphics[width=0.49\columnwidth]{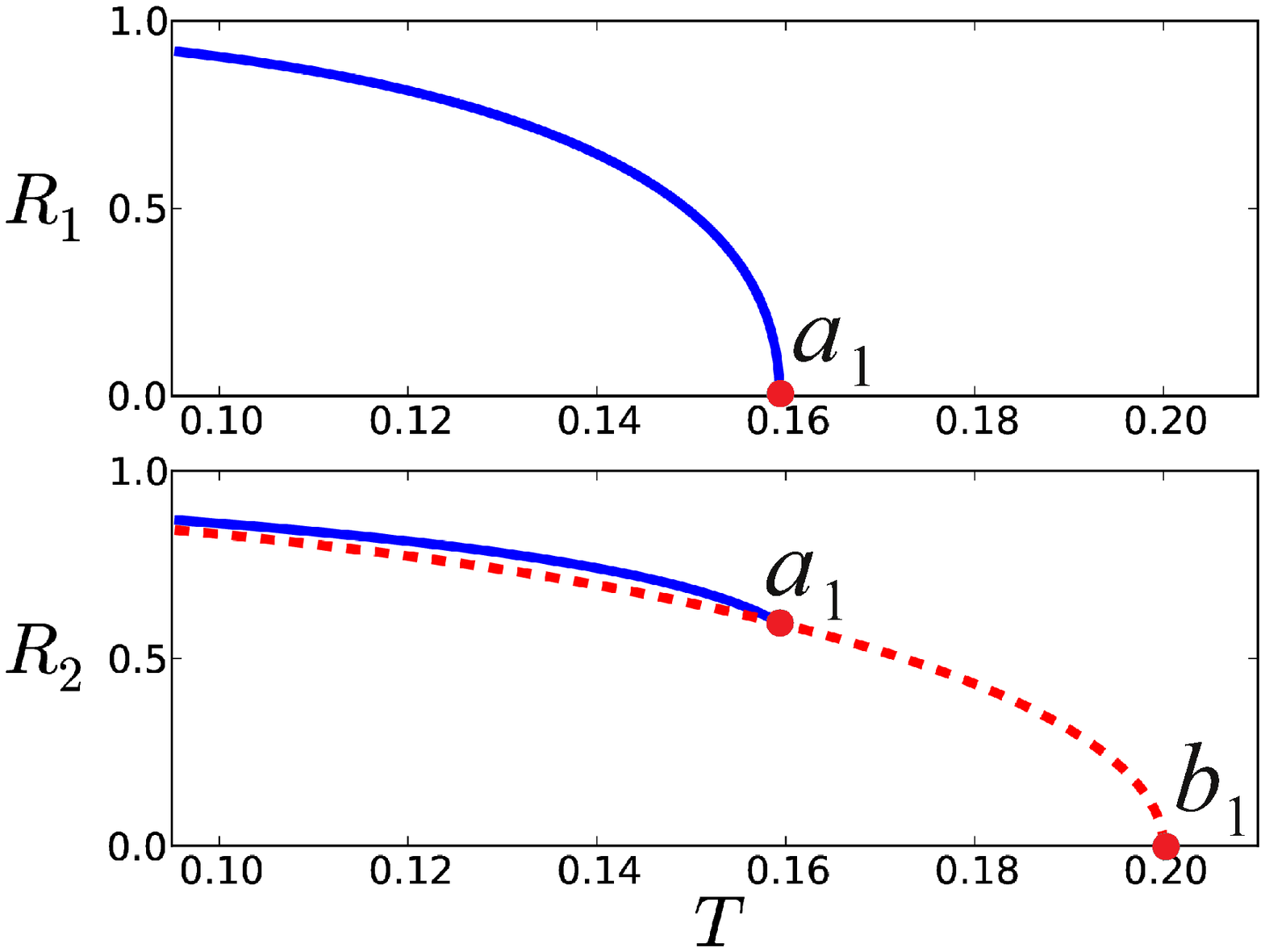}}
		\caption{Dependencies of order parameters $R_{12}$ on the 
disorder parameter $T$, 
		for $s=1$ and different values of $q$: (a) $q=0.9$, (b)
$q=0.36$, (c) $q=0.28$, (d) $q = 0.2$. Solid blue line:
		branch of general solution $R_1\neq 0$; dashed red line: branch
of the two-cluster state with $R_1=0$, $R_2\neq 0$.}
		\label{fig:r12_s1}
	\end{figure}

We present the diagram of synchronous states in the parameter plane $(q,T)$, for the
case of identical oscillators $s=1$
in the~ Fig. \ref{fig:qt_id}. 
Here also the stability lines of the disordered state
$\rho=(2\pi)^{-1}$ are shown with dashed 
lines (see Appendix~\ref{App:Asynch.stab} for details of calculation). Three
major states can be observed: a disordered one, one with all non-zero
order parameters, and a two-cluster one where all odd order parameters vanish.
For small and large values of parameter $q$, i.e.
where one of the coupling modes dominates (the first harmonics coupling for $q$
close to one or the second harmonics
coupling for $q$ close to zero) the transitions are supercritical.
We illustrate them in Fig.~\ref{fig:r12_s1} (panels (a),(d)), showing
dependencies of the order parameters on $T$.
In the middle part of the phase diagram (between the points marked $p_1$,$p_2$
in Fig.~\ref{fig:qt_id}), for $q$ close to $0.3$, the transitions are
subcritical,
so that a bistability occurs. These regimes are illustrated in
Fig.~\ref{fig:r12_s1} (panels (b),(c)). Transition 
from the disordered to the two-cluster state is always supercritical (see dashed
red line in panels (c,d) of Fig.~\ref{fig:r12_s1}).

\subsection{General phase diagram}

\begin{figure}[!htb]	
\centerline{(a)\includegraphics[width=0.49\columnwidth]{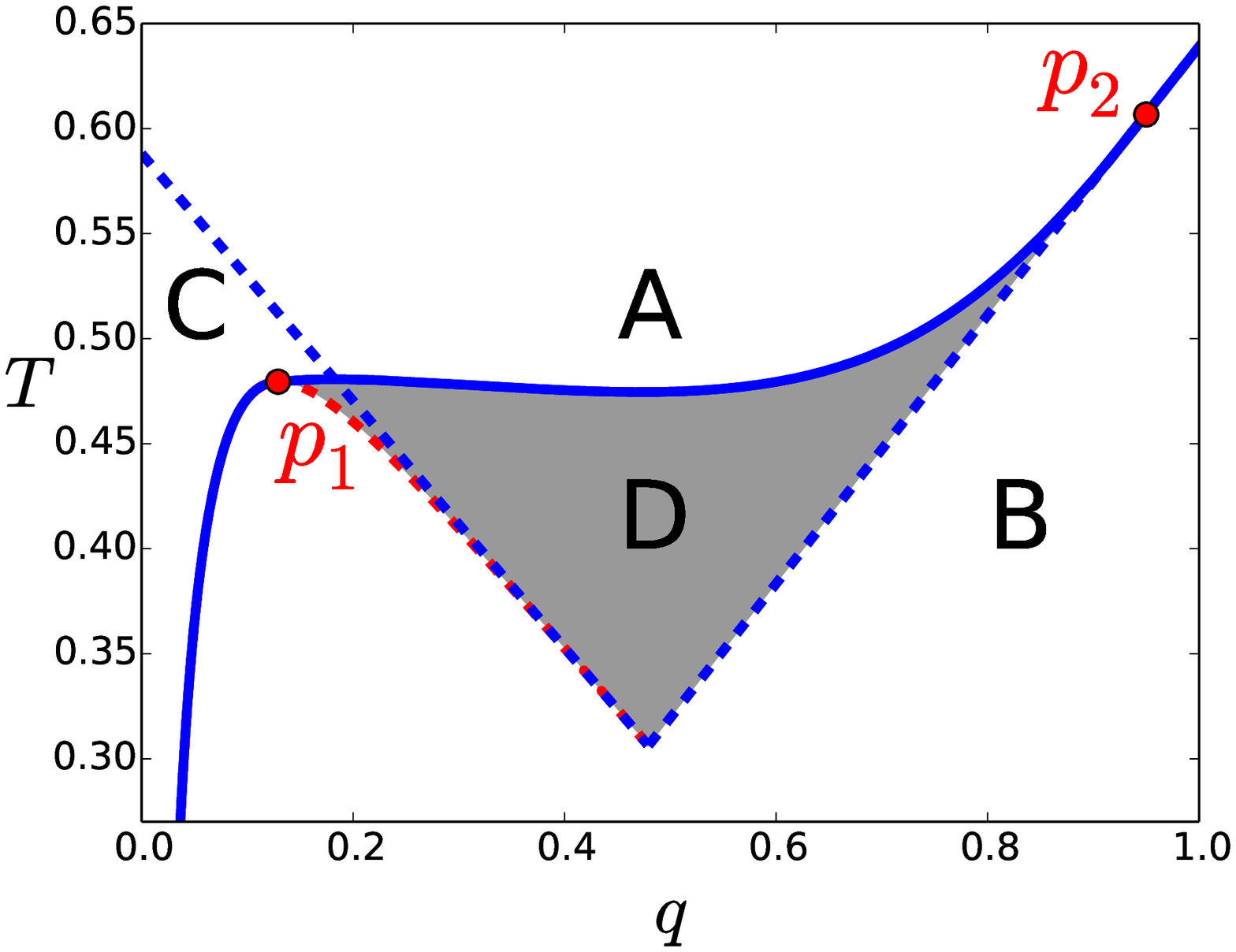}
(b)\includegraphics[width=0.49\columnwidth]{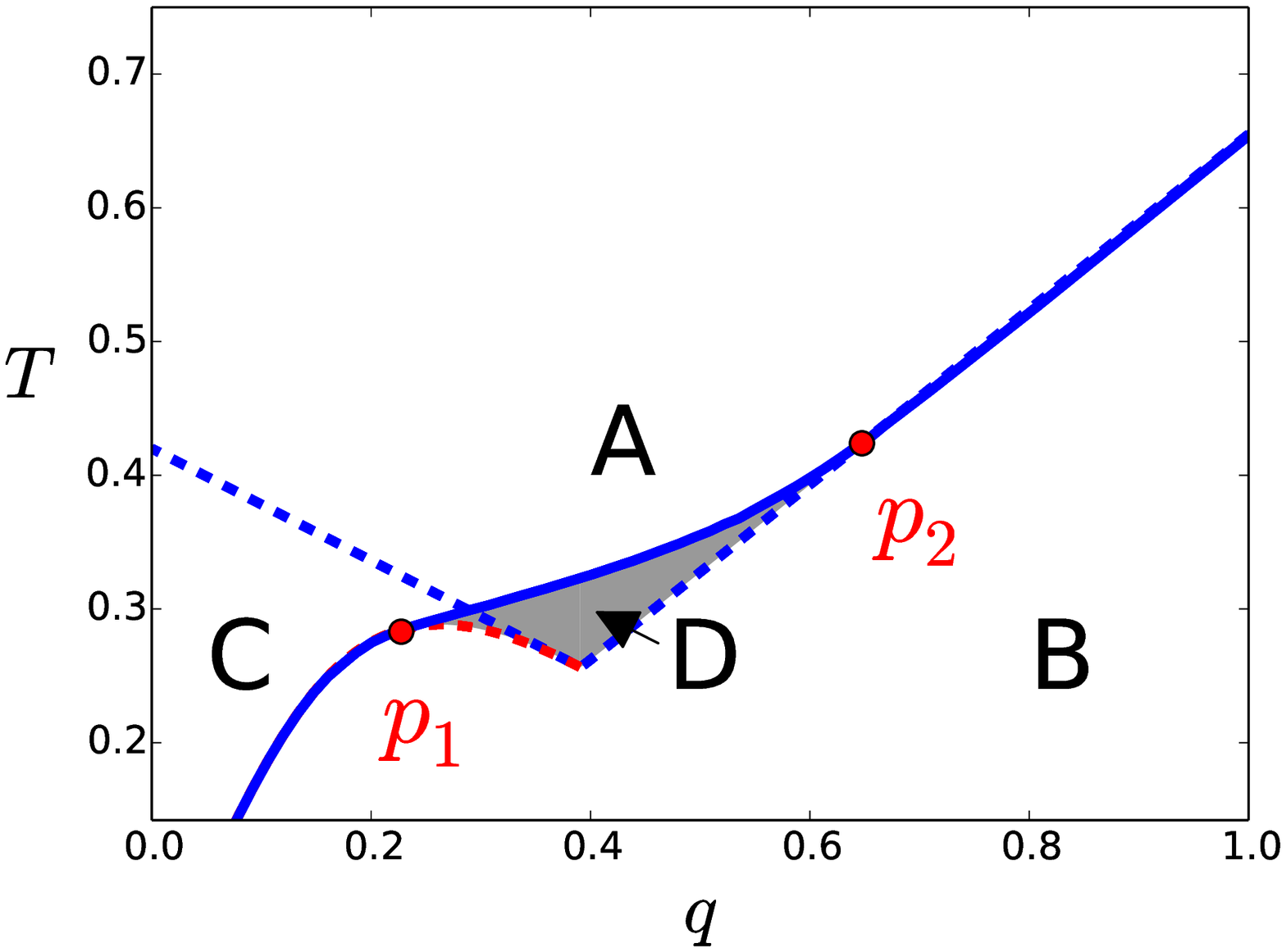}}
		\caption{The same as Fig.~\ref{fig:qt_id}, but for $s=0.1$ (a)
and $s=0.5$ (b).
	Region E is not denoted because it is very tine on these plots.}
		\label{fig:qt_s01_s09}
	\end{figure}

The phase diagrams on plane of basic
parameters	$(q,T)$ for other values of $s$ are qualitatively the same
as Fig.~\ref{fig:qt_id}.  We show two cases $s=0.1$ and $s=0.5$ 
in Fig.~\ref{fig:qt_s01_s09}.
All these diagrams are qualitatively similar, and we expect also that
stability properties of different solutions are like in Fig.~\ref{fig:r12_s1}.

\section{Limiting case of small noise}
In this section we describe the limit of small noise $s\to 0$, that corresponds
to the noise-free case of bi-harmonic Kuramoto model studied in
\cite{Daido-95,Daido-96a,Chiba-Nishikawa-11,Komarov-Pikovsky-13a,
Komarov-Pikovsky-14,Li-Ma-Li-Yang-14}. This is the mostly challenging part, both
numerically and theoretically.
Numerically, there is a problem in finding a stationary probability density from
\eqref{FP.st.2}, because this
density becomes close to a delta-function, i.e. first a large number of Fourier
modes is needed, and second the methods
of solving an algebraic system for these modes converge very slowly.
Theoretically, it is known that for a noise-free case,
there is a multiplicity of solutions coexisting with a neutrally stable
disordered state \cite{Komarov-Pikovsky-13a}, this
degeneracy is lifted by adding an arbitrary small noise. The reason for this
singular limit is that the Fokker-Planck equation behaves singularly
in the limit of small noise: while for any noise the stationary distribution is 
unique, in the noise-free case the order of equation is reduced and there are
multiple solutions of the resulting Liouville equation. 

\begin{figure}
\centering	
(a)\includegraphics[width=0.46\columnwidth]{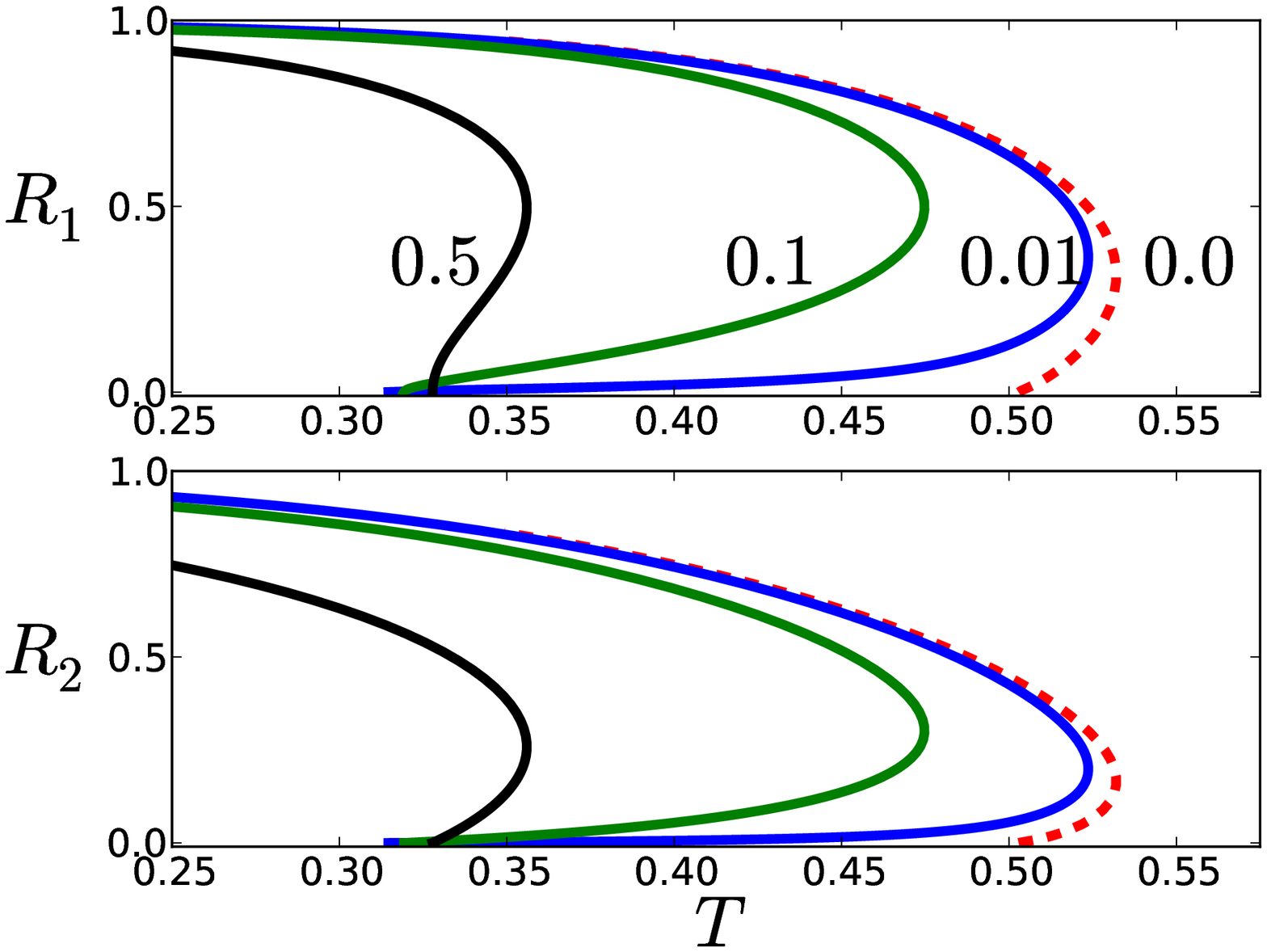}
(b)\includegraphics[width=0.46\columnwidth]{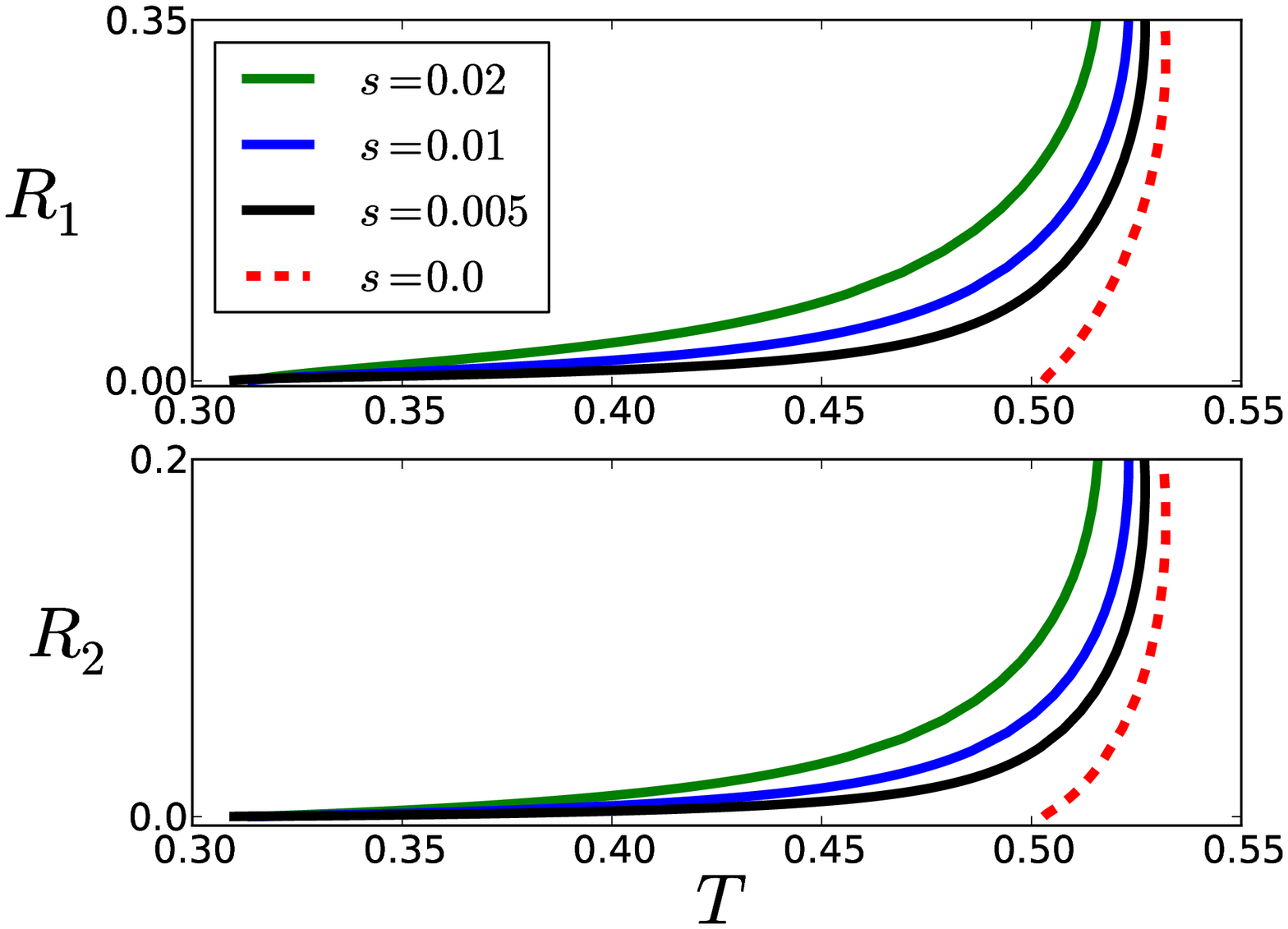}
\caption{(a) The dependencies of order parameters $R_{1,2}$ on
the overall disorder $T$ for $q=0.5$ and different values of $s$. The numbers on
the upper panel depict values of $s$: from the left curve to the right one $s$
changes from 0.5 to zero. The dashed curve corresponds to the special case of
$s=0.0$. 
		(b) The same as in (a) but now curves are plotted near the
points of transition from incoherent to synchronous solutions ($R\to 0$).
}
		\label{fig:r12_s0}
	\end{figure}

In order to shed light on the limit $s\to 0$, we fixed $q=0.5$, and calculated
dependencies $R_{1,2}(T)$ for $s=0$ using the approach
described in \cite{Komarov-Pikovsky-14} for the noise-free case, and the
approach above for different values of parameter $s$.
The results presented in Fig.~\ref{fig:r12_s0} show how the noise-free curve
(dashed red curve in fig.~\ref{fig:r12_s0}(a))
is approached as $s\to 0$ (solid curves in Fig.~\ref{fig:r12_s0}). One can see
that there is a jump in the values of $T$ at which
the curves touch the line $R_{1,2}=0$. For the case of noisy oscillators, these
points are those where
the disordered state loses its stability. For the noise-free case, it is another
point, not related to a stability exchange.
This jump is responsible for a formation of a ``boundary layer", depicted in
more details in Fig.~\ref{fig:r12_s0}(b),
where the noisy curves rapidly escape the vicinity of the curve $s=0$ as
$R_{1,2}$ go to zero.
It appears that the solution of the limiting case $s=0$ always differs
significantly from the noisy solutions $s\neq 0$ at $R\to 0$, regardless of the
noise strength represented by the parameter~$s$.

In the other words, the limit $s\to 0$ for the noisy case does not coincide with
the noise-free solution $s=0$ near the bifurcation point $R=0$.
This effect was also discussed in the papers \cite{Komarov-Pikovsky-13a,
Komarov-Pikovsky-14}.
In the limiting noise-free case $s=0$, the stationary solutions for distribution
function $\rho(\varphi,\omega)$ are always singular: they contain combination of
delta-functions for any small $R_{1,2}$ (see \cite{Komarov-Pikovsky-14} for
detail). 
In contradistinction, the presence of the noise always regularizes solutions
causing smooth 
and non-singular stationary distribution functions $\rho(\varphi,\omega)$.
It is important to notice, that the effective noise appears as a combination
$s/R$ in the algebraic equations 
for stationary modes (\ref{FP.st.Four.2}).
Therefore, for any $s\neq 0$  we always have the limit of effectively ``large
noise'' at the bifurcation point $R\to 0$. 
The latter causes qualitative difference near the transition point $R\to 0$
between noisy ($s\neq 0$) and noise-free ($s=0$) cases even for infinitely small
noise strength.
With decrease of $s$,  the ``boundary layer'' shrinks (the smaller is parameter
$s$, 
the smaller values $R$ we need for effective noise to be large) as one can see
from Fig.~\ref{fig:r12_s0}.

\begin{figure}
\centering	
\psfrag{xlabel}[cc]{$\log_{10} N$}
\psfrag{ylabel}[cc]{$\langle \log_{10} \tau\rangle$}		
\includegraphics[width=0.49\columnwidth]{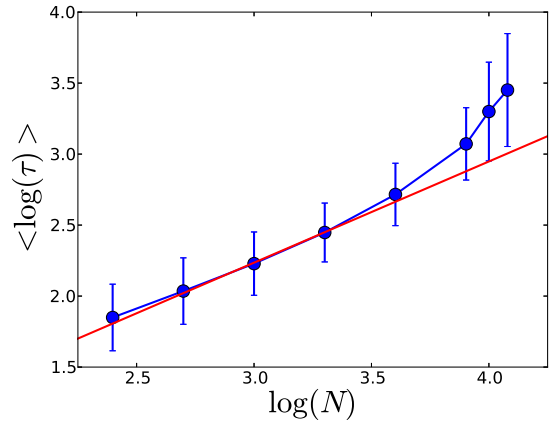}
		\caption{Statistics of the transition time from the incoherent state
(linearly stable in thermodynamic limit) to synchronous mode for the finite
size ensemble. Parameters: $q=0.5$, $T=0.47$, $s=0.01$.}
		\label{fig:times}
	\end{figure}

Remarkably, the incoherent state which is linearly stable
in the thermodynamic limit, can become metastable for
finite-size ensembles in the region of bistability of asynchronous and
synchronous solutions. As one can see from the figure~\ref{fig:r12_s0}(a,b), the lower branch of
synchronous solutions (the unstable one) is relatively close to the incoherent
state $R_{1,2}=0$ at the point $T=0.47$. Therefore, due to the random
finite-size fluctuations, there is always a probability for the system to escape
the small basin of attraction of the disordered state, and to make a transition
to synchronous state which has a much larger basin.
The figure~\ref{fig:times} shows statistics of the life times $\tau$ of the incoherent
solution (linearly stable in thermodynamical limit) for relatively small noise
strength $s=0.01$ and $T=0.47$. For each value of $N$ we performed many runs,
from which the mean value $\langle\log \tau\rangle$ and the standard deviation of
$\log \tau$  has been calculated (so that the errorbars in Fig.~\ref{fig:times}
are not due to insufficient averaging but represent the variability of the life times). 

The averaged time of transition increases as number of
oscillators in the system grows. Remarkably, one can see on
Fig.~\ref{fig:times} a crossover from a regime of a power-law dependence
of the life time on the system size (for $N\lesssim4000$ a relation $\tau\sim N^{0.71}$
holds) to a more rapid decrease of the life time for $N\gtrsim 8000$. We have to compare this
with the noise-free case, where according to~\cite{Komarov-Pikovsky-14} the life time
scales as $\sim N^{0.72}$. One can see that for small system size, i.e. for large
finite-size fluctuations, the noise-free situation is reproduced. We explain this as 
following: small noise makes the asynchronous state only weakly stable, compared to
neutral stability of the noise-free case, and for large finite-size fluctuations
these weak stability is not seen. For smaller fluctuations, the system starts to see
the ``potential barrier'' to overcome to go into the stable state with large order parameter.
Now the life time is the activation time which can be expected to follow an exponential
Kramers' law for very small fluctuations (regime, not accessible for moderate values of $N$
that we are able to study numerically). This explains the crossover observed in Fig.~\ref{fig:times}.

\section{Conclusion}
The paper is devoted to the investigation of the Kuramoto model of globally
coupled oscillators with bi-harmonic coupling function. 
In particular, here we concentrated on the effects caused by the action of the
independent additive white noise forces on the oscillators.
In the first part of the paper we have formulated the self-consistent theory
that allows one to find stationary solutions for the system in the
thermodynamic limit (when number of oscillators goes to infinity) in the
presence of noise and an arbitrary distribution of frequencies.
As a result of the developed theory, we calculated general bifurcation diagram
and described all possible stationary regimes of the model, for different values
of noise strength, spread of frequencies distribution, and coupling constants.

In the noise-free case, as has been shown in
\cite{Komarov-Pikovsky-13a,Komarov-Pikovsky-14}, there exists a multiplicity
(multistability) of the coherent states due to the presence of the second
harmonics in the coupling.
In this work we have shown that the action of white noise withdraws the
multiplicity, however the noise causes several additional complications to the
bifurcation diagram.

First of all, due to the noise, the model contains large area of parameters with
so-called ``nematic phase'',  which represents synchronous two-cluster state
with zero first order parameter ($R_1=0$, $R_2\neq 0$).
Depending on the parameter values, the transitions to synchrony can be
complicated, where possible scenarios include:
\begin{itemize}
\item a simple supercritical (second-order) transition to synchrony, where both
order parameters scale as a square root of supercriticality. This case is
similar to the standard Kuramoto model with pure sinusoidal coupling and occurs
when the second harmonics coupling is relatively small;
\item a subcritical (first-order) transition to synchrony, with a large area of
bistability of coherent and asynchronous solutions. This type of transition
becomes dominating as noise strength goes to zero and the second harmonics in
the coupling function remains relatively strong;
\item first a two-cluster state appears via a supercritical (second-order)
transition. This state is characterized by zero first order parameter and
square-root scaling of the second order parameter. As disorder decreases,
a general synchronous state appears via a subcritical (first-order) transition. 
This happens in the case of relatively weak first harmonic in the coupling
function;
\item first a two-cluster state appears via a supercritical (second-order)
transition. As disorder decreases,
a general synchronous state appears via a supercritical (second-order)
transition, at which the amplitude of the first order parameter grows
continuously. 
This happens in the case of very small first harmonics term in the coupling
function.
\end{itemize}

We also report on the finite-size-induced transitions to synchrony for finite 
ensembles. 
As it was mentioned before, in the thermodynamic limit the transition to
synchrony can be hysteretic with large area of bistability of synchronous and
asynchronous solutions when the second harmonic is relatively large.
In the latter case, noise causes a transition from incoherent state (linearly
stable in the thermodynamic limit) to synchronous solution for finite-size
ensembles.

Finally, we would like to mention an analogy of our model of coupled oscillators
with bi-harmonic coupling to a popular in statistical mechanics XY-model of 
globally coupled spins with ferromagnetic and nematic 
coupling~\cite{Teles_etal-12,Pikovsky-Gupta-et-al-14}. In the latter context
the desynchronized state is a disordered one, while one-cluster and two-cluster states
are ferromagnetic and nematic ones. Two main differences are: (i) in the context of
statistical mechanics, stability of solutions is established via the minimization of
free energy in the canonical description or maximization of entropy in the microcanonical one,
while in the dynamical fomulation above stability properties are defined locally;
(ii) diversity of oscillators' frequencies is a non-equilibrium feature not appearing
in the equilibrium formulation of the XY model (cf.~\cite{Komarov-Gupta-Pikovsky-14} for a 
comparison of the
Kuramoto model with corresponding models from statistical physics).

\begin{acknowledgments}
V.V. thanks the IRTG 1740/TRP 2011/50151-0, funded
by the DFG/FAPESP. M. K. thanks Alexander von Humboldt foundation and the Russian Science Foundation (Project No. 14-12-00811) for support.
A. P. acknowledges the Galileo Galilei Institute 
for Theoretical Physics (Florence, Italy) for the hospitality and the INFN for
partial 
support during the completion of part of this work, and supported by 
the grant (agreement 02.В.49.21.0003 of August 27, 2013  between the 
Russian Ministry of Education and Science and Lobachevsky State University of
Nizhni Novgorod)
\end{acknowledgments}


	
\appendix\section{Stability analysis of the incoherent solution}
\label{App:Asynch.stab}
	
	The detail stability analysis of the incoherent solution of the system
of phase equations with multi-harmonic coupling function has been performed
in~\cite{Crawford-95,Crawford-Davies-99}.
	Here we will present the analysis in the particular case of bi-harmonic
coupling function in the new parameter plane $(q,T,s)$. 

	Consider the following Fokker-Planck equation~(\ref{FP.1}) for
conditional probability density function $\rho(\vp,t\,|\,\omega)$
	\be
		\label{App:FP.1}
		\frac{\partial \rho}{\partial t}+\frac{\partial}{\partial
\vp}\Bigl[\Bigl(\w(1-s)T+q\,{\rm Im}\left(Z_1 e^{-\ii\vp}\right)+(1-q)\,{\rm
Im}\left(Z_2
e^{-2\ii\vp}\right)\Bigr)\rho\Bigr]=sT\frac{\partial^2\rho}{\partial\vp^2}\,,
	\ee
	where 
	\be
		\label{App:Z_m.1}
		Z_{m}(t)=\int \int  g(\omega)\,\rho(\vp,t\,|\,\omega)e^{\ii
m\vp}d\vp \,d\omega.
	\ee
	Then in the Fourier modes representation
	\be	
		\label{App:Four.1}
		\rho(\vp,t\,|\,\w)=\frac{1}{2\pi}\sum_n C_n(t,\w)e^{\ii
n\vp}\qquad C_n(t,\w)=\int_0^{2\pi}\rho e^{-\ii n\vp}d\vp,\qquad C_0(t,\w)=1
	\ee
	we obtain
	\be
		\label{App:FP.Four.0}
		\begin{split}		
\frac{dC_k}{dt}&=\int_0^{2\pi}\Biggl[-\frac{\partial}{\partial
\vp}\Bigl[\Bigl(\w(1-s)T+q\,{\rm Im}\left(Z_1 e^{-\ii\vp}\right)+(1-q)\,{\rm
Im}\left(Z_2
e^{-2\ii\vp}\right)\Bigr)\rho\Bigr]+sT\frac{\partial^2\rho}{\partial\vp^2}\Biggr
] e^{-\ii k\vp}\,d\vp=\\
			&=\Bigl(-\ii k\w(1-s)T-k^2sT\Bigr)C_k+\ii
kq\,\frac{Z_1^*C_{k-1}-Z_1C_{k+1}}{2\ii}+\ii
k(1-q)\frac{Z_2^*C_{k-2}-Z_2C_{k+2}}{2\ii}.
		\end{split}
	\ee
	Thus we obtain the following system for $C_k$
	\be
		\label{App:FP.Four.1}	
\frac{dC_k}{dt}=k\left[-\Bigl(\ii\w(1-s)T+ksT\Bigr)C_k+q\,\frac{Z_1^*C_{k-1}
-Z_1C_{k+1}}{2}+(1-q)\frac{Z_2^*C_{k-2}-Z_2C_{k+2}}{2}\right],
	\ee
	where
	\be
		\label{App:Z_m.2}
		Z_{1,2}(t)=\int  g(\omega)\,C_{1,2}^*(t,\w) \,d\omega.
	\ee
	
	Small perturbation to incoherent solution
$\rho(\vp,t\,|\,\omega)=(2\pi)^{-1}$ means that $C_k<< 1$ for $k\neq 0$ and thus
$Z_m<<1$, so in order to linearize the system~(\ref{App:FP.Four.1}) we can
neglect all the terms such as $C_mC_n$ and $Z_mC_n$ and higher. Then for $k>0$
	\be
		\label{App:FP.Four.lin.1}
		\begin{split}		
\frac{dC_1}{dt}&=-T\Bigl(\ii\w(1-s)+s\Bigr)C_1+q\,\frac{Z_1^*}{2},\\		
\frac{dC_2}{dt}&=-2T\Bigl(\ii\w(1-s)+2s\Bigr)C_2+(1-q)Z_2^*,\\
			\frac{dC_k}{dt}&=-kT\Bigl(\ii\w(1-s)+ks\Bigr)C_k, \ \
\text{for} \ k=3,4,...
		\end{split}
	\ee
	and complex conjugate equations for $k<0$ because $C_{-k}=C^*_k$.
	Introduction to~(\ref{App:FP.Four.lin.1}) of the
expressions~(\ref{App:Z_m.2}) for $Z_{1,2}$ gives integro-differential equations
for $C_{1,2}$
	\be
		\label{App:FP.Four.lin.2}
		\begin{split}		
\frac{dC_1}{dt}&=-T\Bigl(\ii\w(1-s)+s\Bigr)C_1+q\,\frac{1}{2}\int 
g(\omega)\,C_{1}\,d\omega,\\		
\frac{dC_2}{dt}&=-2T\Bigl(\ii\w(1-s)+2s\Bigr)C_2+(1-q)\int  g(\omega)\,C_{2}
\,d\omega,\\
			\frac{dC_k}{dt}&=-kT\Bigl(\ii\w(1-s)+ks\Bigr)C_k, \ \
\text{for} \ k=3,4,...
		\end{split}
	\ee
	From~(\ref{App:FP.Four.lin.2}) follows that the equations for the
harmonics split and because $T\ge0$ and $0\le s\le1$ all the high harmonics with
$k\ge3$ and their complex conjugates are stable, whereas instability appears in
first and second harmonics independently, depending on $q,T$.
	
	Since the equations for modes~(\ref{App:FP.Four.lin.2}) are decoupled in
order to find boundary conditions when the first and second harmonics become
unstable one should put in~(\ref {App:FP.Four.lin.2}) $dC_1/dt=0$ and
$dC_2/dt=0$ and self-consistently obtain two conditions on the parameters $q,
T$. Then, by using expression~(\ref{App:Z_m.2}) we obtain
	\be
		\label{App:s-f.FP.Four.lin}
		\begin{split}
			T\Bigl(\ii\w(1-s)+s\Bigr)C_1&=q\,\frac{1}{2}\,Z_1^*,\\
			2T\Bigl(\ii\w(1-s)+2s\Bigr)C_2&=(1-q)\,Z_2^*.
		\end{split}
	\ee
	Introducing~(\ref{App:s-f.FP.Four.lin}) to~(\ref{App:Z_m.2})
	\be
		\label{App:s-f.Z_m.1}
		\begin{split}
			Z_1^*&=\frac{q}{2T}\int
\frac{g(\w)\,Z_1^*}{\ii\w(1-s)+s}\,d\w,\\
			Z_2^*&=\frac{1-q}{2T}\int
\frac{g(\w)\,Z_2^*}{\ii\w(1-s)+2s}\,d\w,
		\end{split}
	\ee
	we obtain two lines on the $(q,T)$ parameter plane for any given $s$:
	\be
		\label{App:qT.asynch.R1}
		T=q\,{1\over 2}\int\frac{g(\w)s}{\w^2(1-s)^2+s^2}\,d\w,
	\ee
	and
	\be
		\label{App:qT.asynch.R2}
		T=(1-q)\int\frac{g(\w)s}{\w^2(1-s)^2+4s^2}\,d\w.
	\ee
	Where the line~(\ref{App:qT.asynch.R1}) on the $(q,T)$ plane corresponds
to the linear stability boundary for $R_1=0$ and another
line~(\ref{App:qT.asynch.R2}) corresponds to the linear stability boundary for
$R_2=0$.

%

\end{document}